\documentclass[aps,prb,twocolumn,showpacs,superscriptaddress,10pt]{revtex4-1}\begin{tiny}\end{tiny}

\usepackage{longtable}
\usepackage{amsfonts}
\usepackage[dvipdfmx]{graphicx}
\usepackage{rotating}
\usepackage{hyperref}
\usepackage{dcolumn}
\newcolumntype{d}{D{.}{.}{3.5}}
\usepackage[usenames, dvipsnames]{color}

\begin{document}

\title{High-$T_c$ superconductivity in weakly electron-doped HfNCl}

\author{Bet\"{u}l Pamuk}
 \affiliation{CNRS, UMR 7590 and Sorbonne Universit\'{e}s, UPMC Universit\'{e} Paris 06, IMPMC - Institut de Min\'{e}ralogie, de Physique des Mat\'{e}riaux, et de Cosmochimie, 4 place Jussieu, F-75005 Paris, France}
 \affiliation{School of Applied and Engineering Physics, Cornell University, Ithaca, New York 14853, USA}

\author{Francesco Mauri}
 \email{francesco.mauri@uniroma1.it}
 \affiliation{Dipartimento di Fisica, Universit\`{a} di Roma La Sapienza, Piazzale Aldo Moro 5, I-00185 Rome, Italy}
 \affiliation{Graphene Labs, Fondazione Istituto Italiano di Tecnologia, Via Morego, I-16163 Genoa, Italy}
 
\author{Matteo Calandra}
 \email{matteo.calandra@upmc.fr}
 \affiliation{CNRS, UMR 7590 and Sorbonne Universit\'{e}s, UPMC Universit\'{e} Paris 06, IMPMC - Institut de Min\'{e}ralogie, de Physique des Mat\'{e}riaux, et de Cosmochimie, 4 place Jussieu, F-75005 Paris, France}

\date{\today}


\begin{abstract}

We investigate the magnetic and superconducting properties in
electron-doped Li$_x$HfNCl. 
HfNCl is a band insulator that undergoes an insulator to superconductor
transition upon doping at $x\approx0.13$.
The persistence of the 
insulating state for $x<0.13$ is due to an Anderson 
transition probably related to Li disorder. 
In the metallic and superconducting phase,
Li$_x$HfNCl is a prototype two-dimensional 
two-valley electron gas with parabolic bands.
By performing a model random phase approximation approach as well as first-principles
range-separated Heyd-Scuseria-Ernzerhof (HSE06) calculations, we find that the spin
susceptibility $\chi_s$ is strongly enhanced
in the low-doping regime by the electron-electron
interaction. Furthermore, in the low-doping limit, the  exchange interaction renormalizes the intervalley
electron-phonon coupling and results
in a strong increase of the superconducting critical
temperature for  $x<0.15$. On the contrary,  for $x>0.15$, $T_c$ is approximately constant, in
agreement with experiments.
At $x=0.055$ we found that $T_c$ can be as large as 40 K, suggesting
that the synthesis of cleaner samples of Li$_x$HfNCl 
could remove the Anderson insulating state competing with
superconductivity and generate a high-$T_c$ superconductor.

\end{abstract}


\maketitle


\section{Introduction}

The low-doping limit of multivalley semiconductors has
recently been proposed as an alternative route
to achieve high-$T_c$ superconductivity
\cite{Saito2015,Ye2015,Kasahara2015}.
Transition metal dichalcogenides \cite{Novoselov2005,Xu2014,Zhang2014,Ye2012},
ternary transition-metal dinitrides \cite{Gregory1998}, and
cloronitrides \cite{Yamanaka1996,Yamanaka1998}
have been reported to achieve fairly high $T_c$ upon doping.
It is possible to dope multivalley semiconductors up to electron densities of 
$n \sim 10^{14}$ cm$^{-2}$ via field-effect doping \cite{Novoselov2005,Xu2014,Ye2010,Kasahara2011,Thomas2014,Saito2015}.
The doping of these materials can be also be achieved and controlled by
intercalation \cite{Yamanaka1996,Yamanaka1998,Taguchi2006,Takano2008,Takano2008b,Yamanaka2009}.
However, reaching the low-doping limit can be
difficult as disorder and the consequent Anderson transition can
suppress superconductivity.

In two-dimensional and quasi-two-dimensional (2D) semiconductors,
in the weakly doped regime, the density of states (DOS) is constant.
This is different from three-dimensional (3D) semiconductors with parabolic bands,
where generally, as the number of electrons increases, 
the density of states increases as $\sqrt{\epsilon_F}$,
$\epsilon_F $ being the Fermi level.
Therefore, in 3D semiconductors, a large number of carriers is needed
\cite{Ekimov2004}
to achieve a sizable density of states at the Fermi 
level $N(0)$.
As in a phonon-mediated mechanism, $T_c\sim N(0)$,
in a 2D semiconductor, $T_c$ is expected to be constant
because of the constant DOS,
as long as the phonon spectrum is weakly affected by doping.
However, in the weakly doped regime of transition-metal chloronitrides,
$T_c$ {\it increases with decreasing} doping \cite{Yamanaka1996,Yamanaka1998,Taguchi2006}.
This unexpected behavior resulted in a search for
a theoretical understanding of the physics of superconductivity in 2D semiconductors
\cite{Weht1999,Heid2005,GalliPickett2010,Akashi2012,Kotliar2013,Botana2014,Paolo2015,Pamuk2016}.

In previous work, it has been shown that in 2D multivalley semiconductors,
at low doping, the electron-electron interaction enhances intervalley 
electron-phonon coupling,
explaining the behavior of $T_c$ \cite{Paolo2015,Pamuk2016}.
The enhancement of $T_c$ is linked to the enhancement 
of the spin susceptibility $\chi_s$.
Furthermore, a systematic study of the electronic, magnetic, and vibrational properties
of Li$_x$ZrNCl has been performed using density functional theory (DFT) with hybrid functionals
with exact exchange and range separation,
and this paper shows that the exact exchange component leads to a similar enhancement in 
spin susceptibility and electron-phonon interaction \cite{Pamuk2016}. 
This effect on the enhancement of $T_c$ should be quite general as it only requires
basic general ingredients such as 
a 2D multivalley (ideally two-valley) semiconductor and
a large enough electron-gas parameter, $r_s=1/a_{\rm B}\sqrt{\pi n}$
with $a_B=\epsilon_M \hbar^2/(m^* e^2)$ where $n$ is the electron
density per unit area 
[linked to the doping per formula unit $x$ per area $\Omega$ of 2 formula units (f.u.) for Li$_x$ZrNCl: $n=2x/\Omega$], 
$\epsilon_M$ is the environmental dielectric
constant (i.e., the dielectric constant of the undoped semiconductor), 
and $m^*$ the effective mass of the electronic band
\cite{Paolo2015}.
Therefore, it is natural to search for high-$T_c$ superconductivity in
other materials with either
larger $\epsilon_M$ or with lower $n$ and $m^*$.

An interesting system with these features can be intercalated HfNCl.
Superconductivity has been observed 
with Li-intercalated $\beta$-HfNCl with $T_c=20$ K \cite{Takano2008},
and with co-intercalated  Li$_{0.48}$(THF)$_y$HfNCl with $T_c=25.5$ K \cite{Yamanaka1998,Takano2008}.
As $\beta$-ZrNCl, $\beta$-HfNCl is a two-dimensional two-valley semiconductor
with an almost perfect parabolic conduction band and constant DOS. 
Moreover, in $\beta$-HfNCl, $\epsilon_M=4.93$ \cite{GalliPickett2010} is slightly smaller than
in the case of $\beta$-ZrNCl ($\epsilon_M=5.59)$.
Thus, it is natural to expect that a similar enhancement in $T_c$ at low doping
occurs also in Li$_x$HfNCl.
However, the $T_c$ in Li$_x$HfNCl is surprisingly flat in the weakly doped regime,
and an Anderson transition occurs at almost three times larger doping ($x\approx0.15$)
with respect to Li$_x$ZrNCl.
It is then possible that the Anderson transition prevents the
enhancement of $T_c$ at low doping, or, alternatively,
the reported doping is indeed nominal doping and not the real electron doping
occurring in the sample.
More experimental insight into the low-doping regime can also be obtained by field-effect doping.
In this paper, we follow the method introduced in Refs. \onlinecite{Paolo2015, Pamuk2016}
to explore the behavior in Li$_x$HfNCl.
We propose that clean samples at sufficiently low doping can achieve 
higher $T_c$ without the need of further cointercalation.

\section{Computational details}

Calculations are performed using the 
{\sc Quantum ESPRESSO} \textit{ab initio} method \cite{QE}
with the generalized gradient approximation (GGA)
as implemented in the Perdew-Burke-Ernzerhof (PBE) functional \cite{PBE} 
with ultrasoft norm conserving pseudopotentials and 
plane wave basis sets.
The doping of the semiconductor is simulated by 
changing the number of electrons and adding a compensating jellium background, 
which has been previously shown to give accurate results \cite{Heid2005,Botana2014}. 
The atomic coordinates are relaxed with lattice parameters fixed at the 
experimental values from Ref. \onlinecite{Takano2008}.
For the energy convergence, a threshold 
on the change in total energy of $10^{-10}$ Ry
is used for all calculations.
A Methfessel-Paxton smearing of 0.01 Ry with
an electron-momentum grid of $48 \times 48 \times 48$
are used for the relaxation of the internal coordinates 
and for calculating the electronic band structure.
The density of states is calculated using a Gaussian smearing of 0.01 Ry.

Furthermore, we have performed calculations with the Heyd-Scuseria-Ernzerhof (HSE06) \cite{HSE06}
functional that has exact exchange and range separation components,
using the CRYSTAL code \cite{Crystal14}
with Gaussian-type triple-$\zeta$ valence polarized basis set orbitals \cite{TZVPbasis, Hfbasis},
where the diffuse Gaussian functions of the Hf basis are reoptimized.
A Fermi-Dirac smearing of 0.0025 Ha,
an electron-momentum grid of $48\times48\times16$,
an energy convergence threshold of $10^{-9}$ Ha,
and real space integration tolerances of 8-12-8-30-60,
with a sixth order multipolar expansion
are used for the HSE06 calculations.

The effective mass $m^*$ is calculated from the curvature of a fourth order polynomial fit to the 
region between the Fermi energy and the conduction band minimum around the special point
\textbf{K}, assuming that the mass tensor is isotropic.

Electron-phonon coupling and phonon frequencies are calculated with
the PBE functional with
a Methfessel-Paxton smearing of 0.02 Ry,
electron-momentum grid of $12\times12\times4$,
Wannierization \cite{Wannier90} of the electronic bands
with an electron-momentum grid of $6\times6\times2$,
correspondingly, a phonon-momentum grid of $6\times6\times2$,
and a Wannier interpolation scheme of electron-phonon coupling with 
a grid of $40\times40\times6$ \cite{WannierInterpol}.

\section{Results and discussion}

\subsection{Electronic structure}

The primitive unit cell of HfNCl has rhombohedral structure (space
group $R\bar{3}m$, No. $166$) with 2 f.u. per unit cell.
It can also be constructed by a conventional cell of hexagonal structure 
with 6 f.u. per cell with ABC stacking.
Instead of using the rhombohedral unit cell, we take advantage of the weak 
interlayer interaction \cite{Kasahara2010,Heid2005,Takano2011,Botana2014, Paolo2015},
which makes the stacking order negligible,
and we adopt a hexagonal HfNCl structure with AAA stacking.
This is equivalent to the hexagonal structure with the space group 
$P\bar{3}m1$ (No. 164),
with 2 f.u. in the unit cell.
We use the experimental lattice parameters $a$ and $c$ for each doping
from Ref. \onlinecite{Takano2008}. 

\begin{figure}[!ht]
        \centering
        \includegraphics[clip=true, trim=-1mm 0mm 0mm 0mm,width=0.45\textwidth]{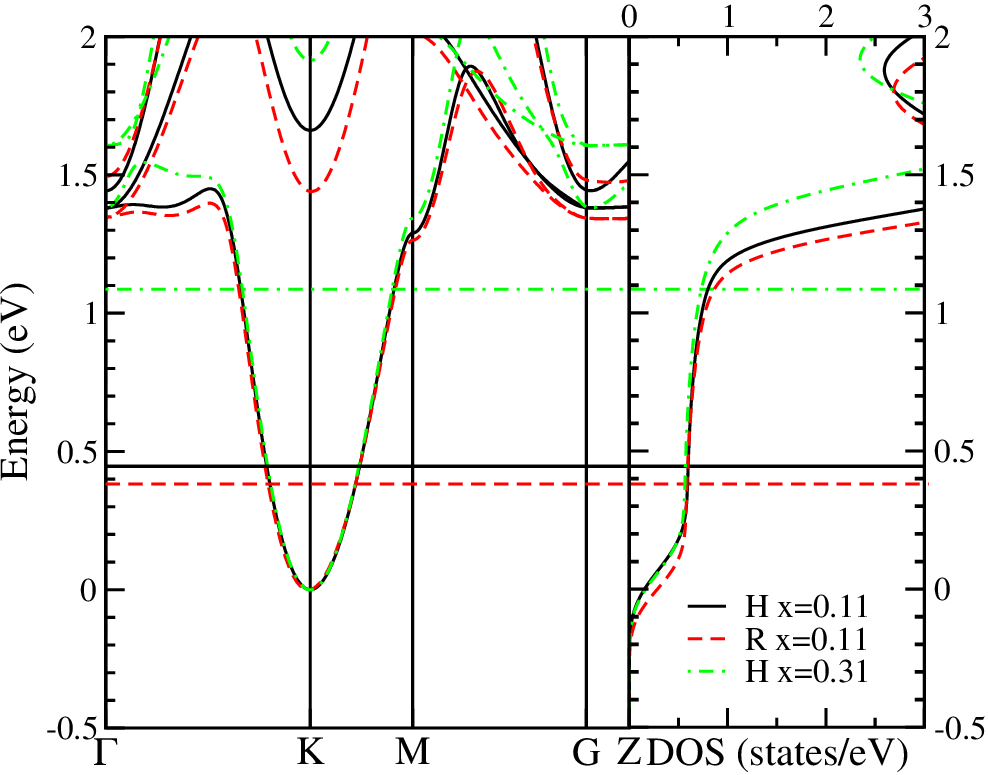}
        \caption{Electronic structure and density of states (DOS) of Li$_x$HfNCl
        calculated with the PBE functional.
        The hexagonal structure (H) with AAA stacking is compared to 
        the rhombohedral structure (R) with ABC stacking
        for the doping $x=0.11$.
        For the hexagonal structure with AAA stacking, the electronic structure of
        the doping $x=0.11$ is compared to that of the doping $x=0.31$.
        The DOS is given in units of states/eV per 2 f.u. of each unit cell.}
        \label{fig:bands}
\end{figure}

To confirm the assumption that the stacking order does not play a 
significant role in the conduction band, 
we compare the electronic bands and the density of states of
hexagonal and rhombohedral structures for the doping $x=0.11$
in Fig. \ref{fig:bands}.
The electronic structure is not
affected by the stacking difference.

This layered system can be considered as the prototype of a 2D two-valley
electron gas. Indeed, the bottom of the  conduction band of HfNCl is
composed of two perfectly parabolic bands at points ${\bf K}$ and
${\bf K^{\prime}}={\bf 2K}$ in the Brillouin zone. 
The conduction band is a simple parabola, with 
a minimum at the \textbf{K} point of the Brillouin zone,
and the density of states is essentially constant along 
the parabolic part of the conduction band.
The curvature of the rhombohedral structure is slightly smaller,
hence the Fermi energy is slightly lower, than the hexagonal structure.
This difference also would lead to a slightly larger effective mass
calculated with the rhombohedral structure.
The rest of the calculations are performed with the hexagonal structure.

Upon Li intercalation, Li atoms are placed between the HfNCl layers.
Li acts as a donor and gives electrons to the Hf-N layers.
The density of states stays almost constant,
as shown in our virtual crystal calculation for Li$_x$HfNCl in Fig. \ref{fig:bands}.
The semiconducting state is lost with doping and superconductivity emerges. 
While it is well established that in Li$_x$ZrNCl the superconducting state is enhanced 
at low doping \cite{Yamanaka1996,Yamanaka1998,Taguchi2006},
there is no evidence of this enhancement in experiments with
Li$_x$HfNCl.

\begin{table}[!htb] \footnotesize
	\caption{The fundamental band gap, $E_g$
	between the valence band maximum at the $\mathbf{\Gamma}$ point 
	and the conduction band minimum at the $\mathbf{K}$ point,
	effective mass, $m^*$,
	and density of states at the Fermi level, $N(0)$
	of each doping calculated with the PBE and HSE06
	exchange and correlation (XC) functionals with and without exact exchange and range separation.}
	\centering
\begin{ruledtabular}
	\begin{tabular}{l l c c c} 
	$x$   & XC & $E_g$ (eV) & $m^*$ (m$_e$) & $N(0)$ (states/eV) \\
	\hline
	0     & PBE   & 2.203 & 0.615 \\
	0.055 & PBE   & 2.195 & 0.599 & 0.587 \\
	0.11  & PBE   & 2.171 & 0.585 & 0.632 \\
	0.13  & PBE   & 2.168 & 0.580 & 0.639 \\
	0.16  & PBE   & 2.164 & 0.572 & 0.654 \\
	0.18  & PBE   & 2.156 & 0.568 & 0.666 \\
	0.20  & PBE   & 2.153 & 0.564 & 0.680 \\
	0.31  & PBE   & 2.130 & 0.540 & 0.833 \\
	\hline
	0     & HSE06 & 3.330 & 0.522 &       \\
	0.055 & HSE06 & 3.240 & 0.496 & 0.511 \\
	0.11  & HSE06 & 3.148 & 0.472 & 0.539 \\
	0.13  & HSE06 & 3.121 & 0.466 & 0.545 \\
	0.16  & HSE06 & 3.084 & 0.456 & 0.556 \\
	0.18  & HSE06 & 3.055 & 0.451 & 0.565 \\
	0.20  & HSE06 & 3.031 & 0.446 & 0.577 \\
	0.31  & HSE06 & 2.908 & 0.425 & 0.723 \\
	\end{tabular}
\end{ruledtabular}
\label{table:meff}
\end{table}

In Table \ref{table:meff}, we present the band gap $E_g$, effective mass $m^*$, and density of states $N(0)$
of each doping with the PBE and HSE06 functionals.
The band gap $E_g$ decreases with increased doping for both functionals.
As the doping increases, the $m^*$ decreases, and this trend is similar in ZrNCl \cite{Pamuk2016}.
However, in general, the effective mass of HfNCl is slightly larger than that of ZrNCl.
Similarly, $N(0)$ is larger in HfNCl than ZrNCl for all doping \cite{Pamuk2016}.

\subsection{Spin susceptibility}

Similar to $T_c$, the magnetic spin susceptibility
is enhanced in Li$_x$ZrNCl at low doping \cite{Kasahara2009,Taguchi2010},
whereas there are no experiments of spin susceptibility 
as a function of doping for Li$_x$HfNCl.
Spin susceptibility is the response of the spin magnetization to an applied magnetic field,
\begin{equation}
\chi_s=\left(\frac{\partial^2 E}{\partial M^2}\right)^{-1},
\label{ChiM}
\end{equation}
where $E$ and $M$ are the total energy and magnetization, respectively.
The noninteracting spin susceptibility $\chi_{0s}$ is obtained
by neglecting the electron-electron interaction of the conducting electrons.
For perfectly parabolic bands, the noninteracting spin susceptibility
is doping independent and equal to
\begin{equation}
\chi_{0s}=\mu_s N(0)=\frac{g_v m^{*}}{\pi \hbar^2},
\end{equation}
where $\mu_s$ is the Bohr magneton, $g_v$ is the valley degeneracy
($2$ in our case), and $m^{*}$ the band effective mass.
We calculate $\chi_{0s}$ from the density of states of the undoped compound,
and by extrapolating $N(0)$ of the desired doping.
Our calculations show that $\chi_{0s}$ is not enhanced at the low-doping limit.
As $N(0)$ is larger in HfNCl, $\chi_{0s}$ is also larger in HfNCl than ZrNCl \cite{Pamuk2016}.

\begin{figure}[!ht]
        \centering
        \includegraphics[clip=true, trim=0mm 0mm 0mm 0mm,width=0.45\textwidth]{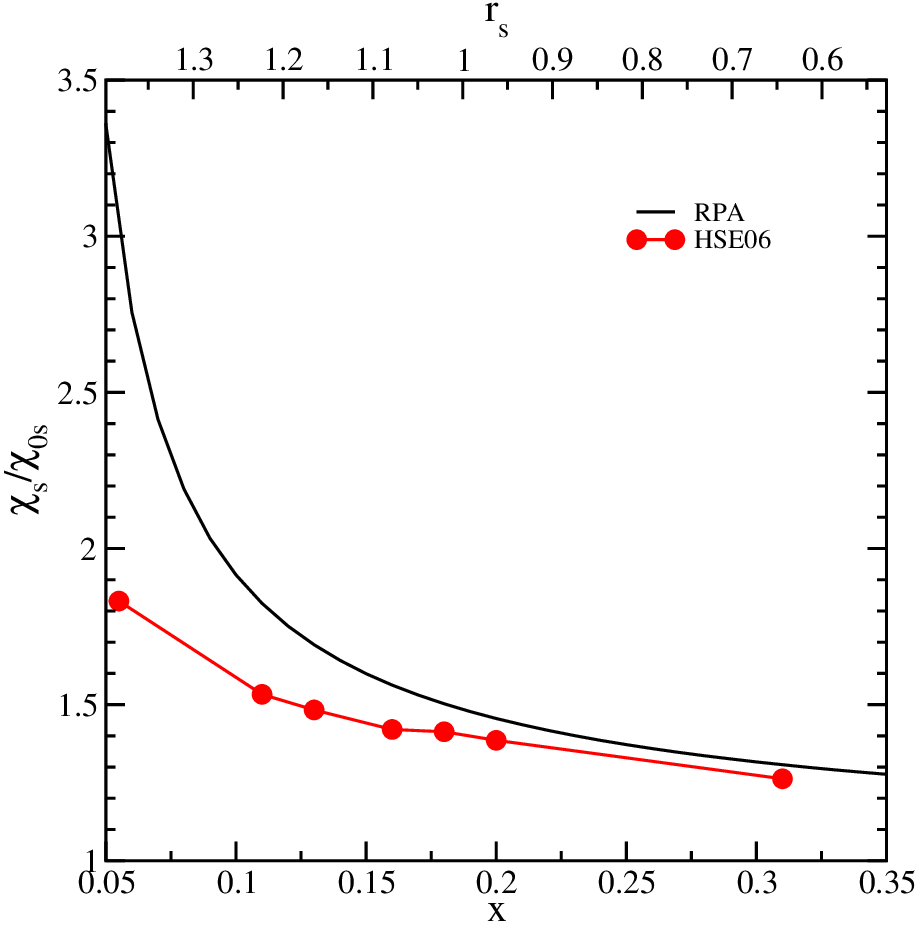}
        \caption{Spin susceptibility enhancement at different doping with
        the RPA and HSE06 approximations.}
        \label{fig:suscep}
\end{figure}

We calculate the spin susceptibility with the HSE06 hybrid functional
by calculating the total energy at fixed magnetization
and then using equation \ref{ChiM} to obtain $\chi_s$. 
We choose the HSE06 functional, because it can reproduce the $\chi_s/\chi_{0s}$ of ZrNCl \cite{Pamuk2016}.
We also compare our results 
with those obtained by a model based on the random
phase approximation (RPA)
\cite{DasSarma2005,Paolo2015}.
The model is appropriate in the low-doping limit where 
$|{\bf k_F}-{\bf K}|<<K$,
a condition necessary to have the intravalley electron-electron scattering
dominating over the intervalley one,
as explained in the Supplemental Material of Ref. \onlinecite{Paolo2015}.
This model assumes a 2D two-valley electron gas with no intervalley Coulomb scattering. 
Therefore, only the intravalley electron-electron
interaction remains and the RPA susceptibility can be calculated
analytically, by using the PBE effective mass of undoped HfNCl
and the environmental dielectric constant $\epsilon_M=4.93$
\cite{GalliPickett2010}.
This value is smaller in HfNCl than ZrNCl ($\epsilon_M=5.59$) \cite{Paolo2015}.

In a 2D two-valley electron gas,
the reduction of doping implies an increase of the $r_s$
electron-gas parameter, 
and, consequently, of the electron-electron interaction \cite{GiulianiVignale}.
The effective mass of $\beta$-HfNCl as calculated by the PBE functional is larger (0.615 m$_e$) than $\beta$-ZrNCl (0.57 m$_e$ \cite{Paolo2015}).
Therefore, both the larger $m^*$ and the smaller $\epsilon_M$ of HfNCl lead to larger $r_s$ as compared to ZrNCl,
at a similar low-doping regime \cite{Paolo2015}.
This implies that the electron-electron interaction is larger in HfNCl,
and hence the spin susceptibility enhancement is also larger in HfNCl.
While the spin susceptibility enhancement at low doping is present for 
both calculations with the RPA and the HSE06 functional,
as presented in Fig. \ref{fig:suscep},
it is milder with the HSE06 functional than the RPA calculation. 

\subsection{Electron-phonon interaction}

The electron-phonon coupling of a mode $\nu$ 
at a phonon momentum \textbf{q}
is defined as
\begin{equation}
{\tilde{\lambda}}_{\mathbf{q}\nu}=\frac{2}{\omega_{\mathbf{q}\nu}^2 N(0) N_k} \sum_k |\tilde{d}^\nu_{\mathbf{k,k+q}}|^2\delta(\epsilon_\mathbf{k})\delta(\epsilon_{\mathbf{k+q}}),
\label{eq:lambda}
\end{equation}
where $\epsilon_{\mathbf{k}}$ is the quasiparticle energy and 
the electron-phonon matrix elements are defined such that
$\tilde{d}^\nu_{\mathbf{k,k+q}}=<\mathbf{k}|\delta\tilde{V}/\delta u_{\mathbf{q}\nu}|\mathbf{k+q}>$,
$u_{\mathbf{q}\nu}$ is the phonon displacement of the mode $\omega_{\mathbf{q}\nu}$,
and $\tilde{V}$ is the single particle potential that is fully screened
by charge, spin, and valley exchange and correlation effects [see
Eq. (2) in Ref. \onlinecite{Paolo2015} for more details].
We first calculate the noninteracting $\lambda_{\mathbf{q}\nu}$ with the PBE functional,
which does not have valley polarization dependence,
using the Wannier interpolation method \cite{WannierInterpol}.

\begin{figure}[!ht]
        \centering
        \includegraphics[clip=true, trim=0mm 0mm 0mm 0mm,width=0.45\textwidth]{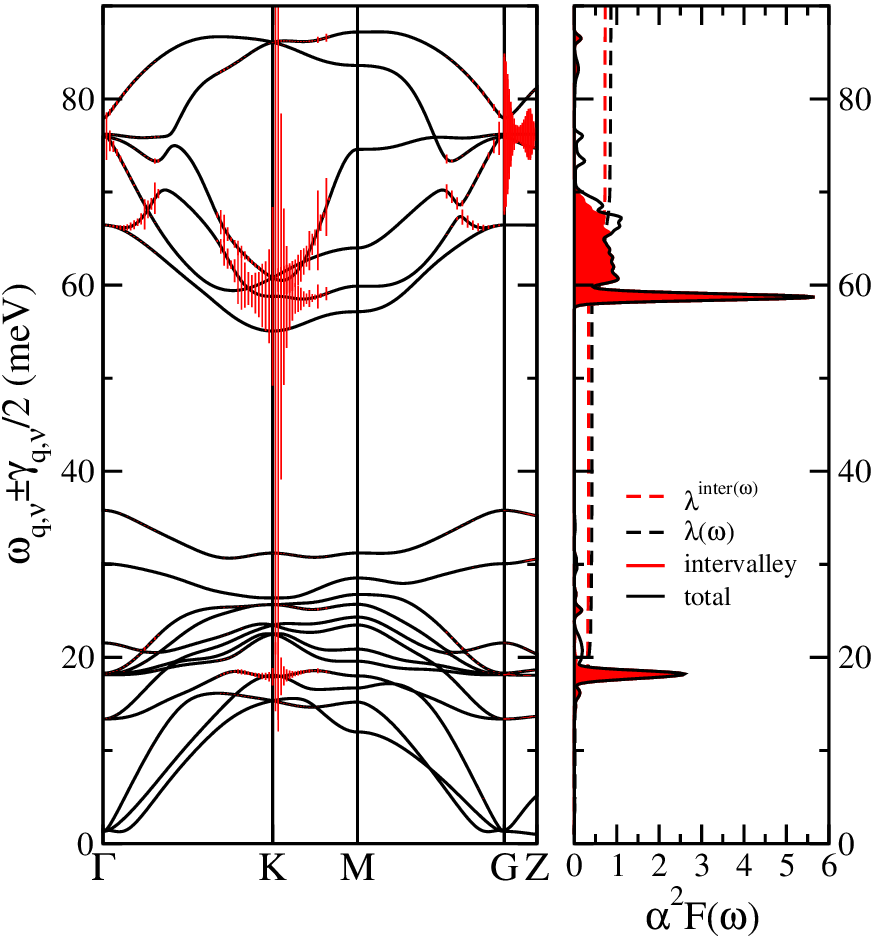}
        \caption{Left: Phonon dispersion along the high symmetry directions of the 
        Brillouin zone for Li$_{0.055}$HfNCl.
		Right: The total and intervalley component of the Eliashberg function $\alpha^2 F(\omega)$
		and the electron-phonon coupling $\lambda(\omega)$.}
        \label{fig:055phon}
\end{figure}

In Fig. \ref{fig:055phon}, we show the phonon dispersion along the high symmetry directions,
and the Eliashberg function $\alpha^2 F(\omega)$ and the electron-phonon coupling $\lambda(\omega)$
for the doping $x=0.055$.
The Eliashberg function has two distinct peaks that are dominated by the modes 
with large phonon linewidths $\gamma_{\mathbf{q}\nu}$
at the \textbf{K} point of the Brillouin zone
at the energies $\sim 19$ meV and $\sim 59$ meV.
To analyze the contribution to the electron-phonon coupling, we separate it into the 
inter- and intra-valley components.
The intervalley electron-phonon coupling $\lambda^{\rm inter}$ is defined such that the modes contributing to the coupling
are in the vicinity of the \textbf{K} and \textbf{2K} points such that, in Eq. (\ref{eq:lambda}),
$\mathbf{k} \in I(\mathbf{K})$ and $\mathbf{k+q} \in I(\mathbf{2K})$; or
$\mathbf{k} \in I(\mathbf{2K})$ and $\mathbf{k+q} \in I(\mathbf{K})$.
The rest of the coupling is attributed to the intravalley electron-phonon coupling $\lambda^{\rm intra}$.
Also shown in Fig. \ref{fig:055phon} is that these modes at the \textbf{K} point
contribute significantly to the intervalley component of the Eliashberg function
and have a large intervalley electron-phonon coupling $\lambda^{\rm inter}$.
Therefore, they induce a valley polarization in this system \cite{Paolo2015}.

Consequently, the spin susceptibility enhancement is directly linked to the enhancement 
in the electron-phonon coupling due to the intervalley interaction \cite{Paolo2015,Pamuk2016}.
The intervalley electron-phonon coupling is enhanced similarly to $\chi_s/\chi_{0s}$
such that
\begin{equation}
	\frac{\tilde{\lambda}^{\rm inter}_{\mathbf{q}\nu}}{\lambda^{\rm inter}_{\mathbf{q}\nu}}=\left(\frac{\chi_s}{\chi_{s0}}\right)^2.
\end{equation}
Following the previously developed methodology, \cite{Paolo2015,Pamuk2016}
we first calculate the bare intervalley electron-phonon coupling $\lambda$ with the PBE functional,
and use the spin susceptibility enhancement of RPA or HSE06
to obtain the corresponding fully dressed coupling $\tilde{\lambda}$.

In Table \ref{table:lambda}, we present the bare electron-phonon coupling $\lambda$,
and its intra- and intervalley components, $\lambda^{\rm intra}$, $\lambda^{\rm inter}$
calculated with the PBE functional,
as well as the fully interacting electron-phonon coupling for the RPA and the HSE06 calculations,
$\tilde{\lambda}^{\rm RPA}$, $\tilde{\lambda}^{\rm HSE06}$ and their corresponding $\omega_{\log}$ values.

\subsection{Superconductivity and $T_c$ enhancement}

\begin{table*}[!htb] \footnotesize
	\caption{For each doping, bare electron-phonon coupling $\lambda$ and its intravalley $\lambda^{\rm intra}$
	and intervalley $\lambda^{\rm inter}$ components as calculated by the PBE functional;
	fully interacting electron-phonon coupling for the RPA and the HSE06 functionals $\tilde{\lambda}^{\rm RPA}$
	and $\tilde{\lambda}^{\rm HSE06}$;
	PBE functional values of $\omega_{\log}^{\rm PBE}$ with intra- and intervalley components
	$\omega_{\log}^{\rm PBE_{intra}}$ and $\omega_{\log}^{\rm PBE_{inter}}$;
	and rescaled $\omega_{\log}^{\rm RPA}$ and $\omega_{\log}^{\rm HSE06}$ in meV.
	The screened Coulomb pseudopotential $\mu^*$ and the $T_c$ values calculated by the RPA and HSE06 functional are also given.
	}
	\centering
\begin{ruledtabular}
	\begin{tabular}{l c c c c c c c c c c c c c} 
		$x$ & $\lambda$ & $\lambda^{\rm intra}$ & $\lambda^{\rm inter}$ & $\tilde{\lambda}^{\rm RPA}$ & $\tilde{\lambda}^{\rm HSE06}$ & $\omega_{\log}^{\rm PBE}$ & $\omega_{\log}^{\rm PBE_{intra}}$ & $\omega_{\log}^{\rm PBE_{inter}}$ & $\omega_{\log}^{\rm RPA}$ & $\omega_{\log}^{\rm HSE06}$ & $\mu^*$ & $T_c^{\rm RPA}$ & $T_c^{\rm HSE06}$ \\
	\hline
	0.055 & 0.861 & 0.133 & 0.728 & 6.730 & 2.578 & 34.219 & 28.494 & 35.385 & 35.233 & 34.991 & 0.326 & 68.60 & 39.39 \\
	0.11  & 0.789 & 0.167 & 0.622 & 2.236 & 1.628 & 32.593 & 28.285 & 33.854 & 33.404 & 33.235 & 0.276 & 38.15 & 25.18 \\
	0.13  & 0.803 & 0.182 & 0.621 & 1.959 & 1.547 & 31.458 & 27.882 & 32.588 & 32.120 & 31.997 & 0.266 & 32.97 & 23.36 \\
	0.16  & 0.860 & 0.208 & 0.652 & 1.800 & 1.523 & 28.643 & 26.451 & 29.380 & 29.026 & 28.962 & 0.254 & 28.17 & 21.86 \\
	0.18  & 0.889 & 0.225 & 0.664 & 1.724 & 1.551 & 26.781 & 25.103 & 27.373 & 27.066 & 27.032 & 0.248 & 25.69 & 21.67 \\
	0.20  & 0.932 & 0.256 & 0.676 & 1.688 & 1.554 & 26.370 & 24.686 & 27.040 & 26.668 & 26.636 & 0.242 & 24.81 & 21.96 \\
	0.31  & 0.973 & 0.372 & 0.601 & 1.401 & 1.331 & 25.915 & 25.956 & 25.887 & 25.905 & 25.906 & 0.222 & 19.89 & 18.30 \\
	\end{tabular}
\end{ruledtabular}
\label{table:lambda}
\end{table*}

Finally, we calculate the superconducting critical temperature $T_c$
using the McMillan-Allen-Dynes equation \cite{McMillan,AllenDynes},
\begin{equation}
T_c=\frac{\omega_{\log}}{1.20} \exp \left(-\frac{1.04(1+\tilde{\lambda})}{\tilde{\lambda} - \mu^* (1+0.62 \tilde{\lambda})} \right),
\label{eq:Tc}
\end{equation}
where $\mu^*=\mu / [1+\mu \log(\epsilon_F/\omega_D)]$ is the screened Coulomb pseudopotential,
with $\epsilon_F$ and $\omega_D=900$ meV being the Fermi and Debye energy, respectively.
We set the unscreened $\mu=0.231$ that gives the correct estimate of the experimental $T_c=19.94$ K
at the highest doping of $x=0.31$ by using the RPA enhanced fully screened 
electron-phonon coupling $\tilde{\lambda}$.
This is in agreement with the GW estimate of $\mu=0.237$ at $x=0.1$ \cite{Akashi2012}.
We present the screened Coulomb pseudopotential $\mu^*$ that is used to calculate $T_c$ for each doping
and the final $T_c$ values for the RPA and HSE06 calculations in Table \ref{table:lambda}.

\begin{figure}[!ht]
        \centering
        \includegraphics[clip=true, trim=0mm 0mm 0mm 0mm,width=0.45\textwidth]{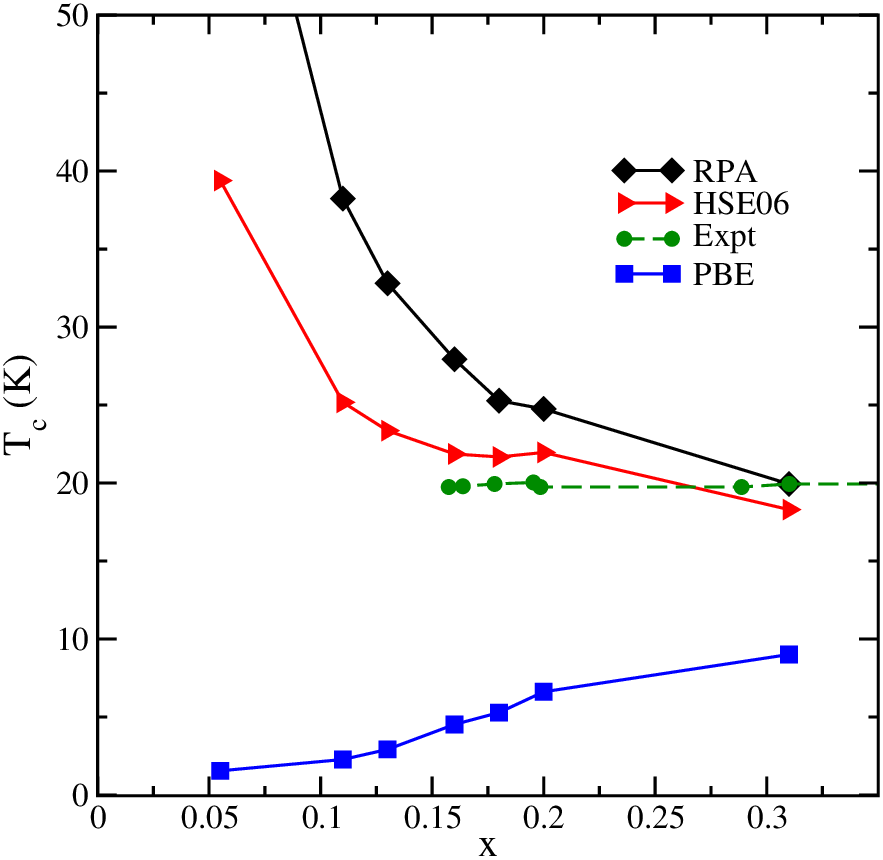}
        \caption{Superconducting critical temperature $T_c$ as a function of doping
        calculated with the bare electron-phonon coupling $\lambda_{\mathbf{q}\nu}$
        as calculated by the PBE functional,
        and with fully dressed electron-phonon coupling, $\tilde{\lambda}_{\mathbf{q}\nu}$
        using the RPA and HSE06 functional.
        The experimental data are taken from Ref. \onlinecite{Takano2008}.
        }
        \label{fig:Tc}
\end{figure}

With the PBE functional, the superconducting temperature, $T_c$ is reduced in
the low-doping limit, in stark disagreement with experiments, 
as shown in Fig. \ref{fig:Tc}.
When the dressing of the intervalley electron-phonon
coupling by the intravalley Coulomb interaction is taken into account,
$T_c$ is enhanced in a similar fashion to what happens to the
spin susceptibility, i.e., it is enhanced significantly, up to $\sim 70$ K, with RPA,
while the enhancement is softer, up to $\sim 40$ K with the HSE06 functional.
In addition, the HSE06 functional agrees well with the experimental $T_c$ for the doping
between $0.15 < x < 0.20$.
We show the details of this scaling for the RPA calculation in Appendix \ref{app:elph},
and we present the phonon dispersion, $\omega$, Eliashberg function $\alpha^2F(\omega)$,
and electron-phonon coupling $\lambda(\omega)$ for the rest of the dopings in Appendix \ref{app:disp}.

\section{Conclusion}

We study the electronic, magnetic, and vibrational properties of Li$_x$HfNCl at the low-doping regime.
We first calculate the electronic structure and find that the effective mass $m^*$ and 
the density of states $N(0)$ are larger in HfNCl as compared to ZrNCl, both for the PBE and the HSE06 functionals.

As there are no experimental data for the spin susceptibility of HfNCl as a function of doping, 
we calculate the spin susceptibility enhancement using both RPA calculations and the HSE06 functional.
Both $m^*$ and $\epsilon_M$ contribute to a larger $r_s$ in HfNCl than ZrNCl.
Therefore, spin susceptibility enhancement is larger in HfNCl than ZrNCl at the low-doping limit,
and this is visible both in the RPA calculations and the HSE06 calculations of $\chi_s/\chi_{0s}$.

Then, we calculate the phonon dispersion $\omega_{\mathbf{q}\nu}$, Eliashberg function $\alpha^2F(\omega)$,
and the bare electron-phonon coupling $\lambda(\omega)$ using the PBE functional.
We further calculate the fully-dressed electron-phonon coupling $\tilde{\lambda}$,
based on the enhancement in the spin susceptibility.

This enhancement is then directly reflected in the calculated $T_c$. 
There is no enhancement in $T_c$ with the PBE functional.
On the other hand, we can speculate that depending on the enhancement in the spin susceptibility, 
high $T_c$ can be reached,
ranging from 40 K (with the HSE06 functional) to 70 K (with the RPA calculation).
Furthermore, the HSE06 functional gives comparable $T_c$ values to the experiments
for dopings $0.15 < x < 0.20$.
However, the $T_c$ goes to zero in experiments for the reported doping $x<0.15$ \cite{Takano2008}.
A possible explanation for this disagreement is that the disorder at the low-doping limit 
can lead to Anderson localization.
Alternatively, it could be that the reported doping is only a nominal doping.
Experiments on field-effect doping can also help one to learn more about the low-doping regime.
In either case, our results predict that the removal of the Anderson transition or
better control of doping in Li$_x$HfNCl could lead to the emergence of a high-$T_c$ superconducting state.

\begin{acknowledgments}
This project has received funding from the Graphene Flagship by the European Union's Horizon 2020 
research and innovation programme under Grant Agreement No. 696656 GrapheneCore1.
We acknowledge PRACE for awarding us access to resource on Marenostrum at BSC based in Spain;
and the computer facilities provided by CINES, IDRIS, CEA TGCC (Grant EDARI No. 2017091202), 
and institute for computing and data sciences (ISCD) at UPMC based in France.
B.P. acknowledges National Science Foundation [Platform for the Accelerated Realization, Analysis, and Discovery of Interface Materials (PARADIM)] under Cooperative Agreement No. DMR-1539918 for her time at Cornell University.
\end{acknowledgments}

\appendix 

\section{Electron-phonon Coupling Scaling}
\label{app:elph}

\begin{figure}[!ht]
        \centering
        \includegraphics[clip=true, trim=0mm 0mm 0mm 0mm,width=0.5\textwidth]{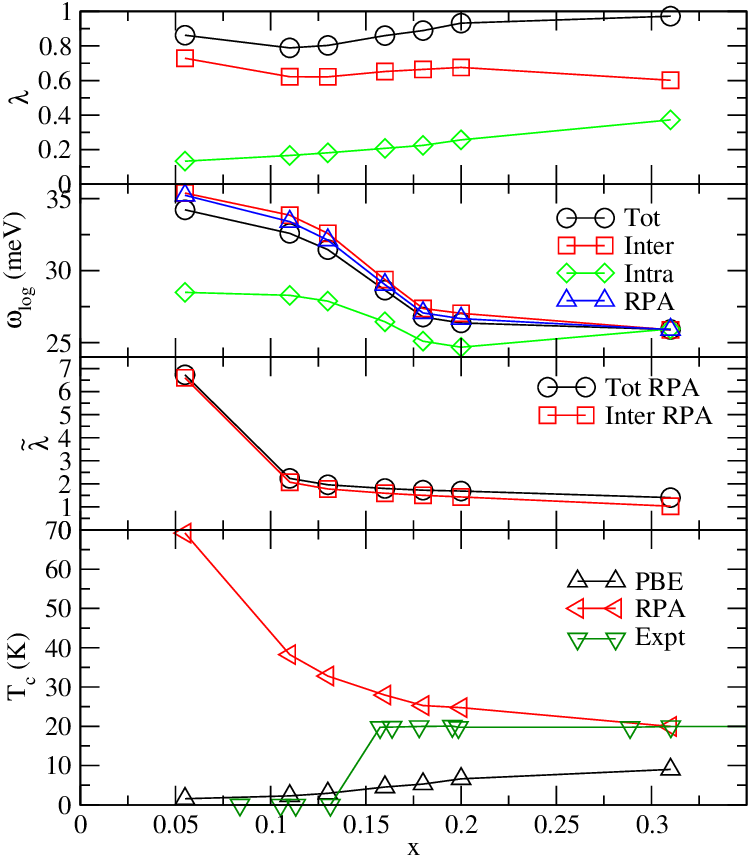}
        \caption{First panel: Average noninteracting
        electron-phonon coupling $\lambda$ for each doping
        including the inter- and intravalley components
        as calculated with the PBE functional.
        Second panel: $\omega_{\rm log}$ for each doping with inter- and intravalley
        components, as well as rescaled $\omega_{\rm log}^{\rm RPA}$.
        Third panel: Interacting electron-phonon coupling $\tilde{\lambda}$ where
        the intervalley term is rescaled with RPA electron-electron interaction enhancement.
        Fourth panel: Superconducting critical temperature $T_c$ as a function of doping,
        calculated by noninteracting (PBE) and interacting (RPA) electron-phonon coupling,
        as compared to the experiments from Ref. \onlinecite{Takano2008}.
        }
        \label{fig:elph}
\end{figure}

We present the electron-phonon coupling in Eq. (\ref{eq:Tc})
as a function of doping in Fig. \ref{fig:elph}.
The top panel shows the average noninteracting electron-phonon coupling $\lambda$,
as well as its inter- and intravalley components.

The second panel shows the total $\omega_{\rm log}$, 
also decomposed into inter- and intravalley components.
In addition, we also present how it is rescaled with the RPA calculation, 
by rescaling $\tilde{\lambda}_{\rm inter}$.
Starting with the definition of $\omega_{\log}$,
\begin{equation}
\omega_{\log}=\exp{\left[ \frac{2}{\lambda} \int_0^{+\infty} \alpha^2 F(\omega) \frac{\log(\omega)}{\omega} d\omega \right]},
\end{equation}
we have separated $\omega_{\log}$ into inter- and intravalley terms.
The intervalley term is
\begin{equation}
\omega_{\log}^{\rm inter}=\exp{\left[ \frac{2}{\lambda^{\rm inter}} \int_0^{+\infty} \alpha^2 F(\omega)^{\rm inter} \frac{\log(\omega)}{\omega} d\omega \right]},
\end{equation}
and the intravalley term is defined similarly.
The relation between these two terms hold such that
\begin{equation}
\omega_{\log}=(\omega_{\log}^{\rm inter}) ^ {\lambda^{\rm inter}/\lambda} \times (\omega_{\log}^{\rm intra}) ^ {\lambda^{\rm intra}/\lambda}.
\label{eq:wlog_tot}
\end{equation}
Therefore, we rescaled it for the RPA calculation
by keeping the intravalley $\lambda^{\rm intra}$ component the same, 
but rescaling the fully interacting intervalley $\tilde{\lambda}^{\rm inter}$
and hence the total $\tilde{\lambda}$ electron-phonon coupling elements.
These are shown in the third panel of the figure for the RPA calculations.

For completeness, we also present the final calculated $T_c$ 
without an intervalley enhancement using the PBE functional,
and with an intervalley enhancement using the RPA calculation, similar to Fig. \ref{fig:Tc}.

\section{Phonon Modes as a Function of Doping}
\label{app:disp}

\begin{figure*}[!ht]
        \centering
        \includegraphics[clip=true, trim=0mm 0mm 0mm 0mm,width=1\textwidth]{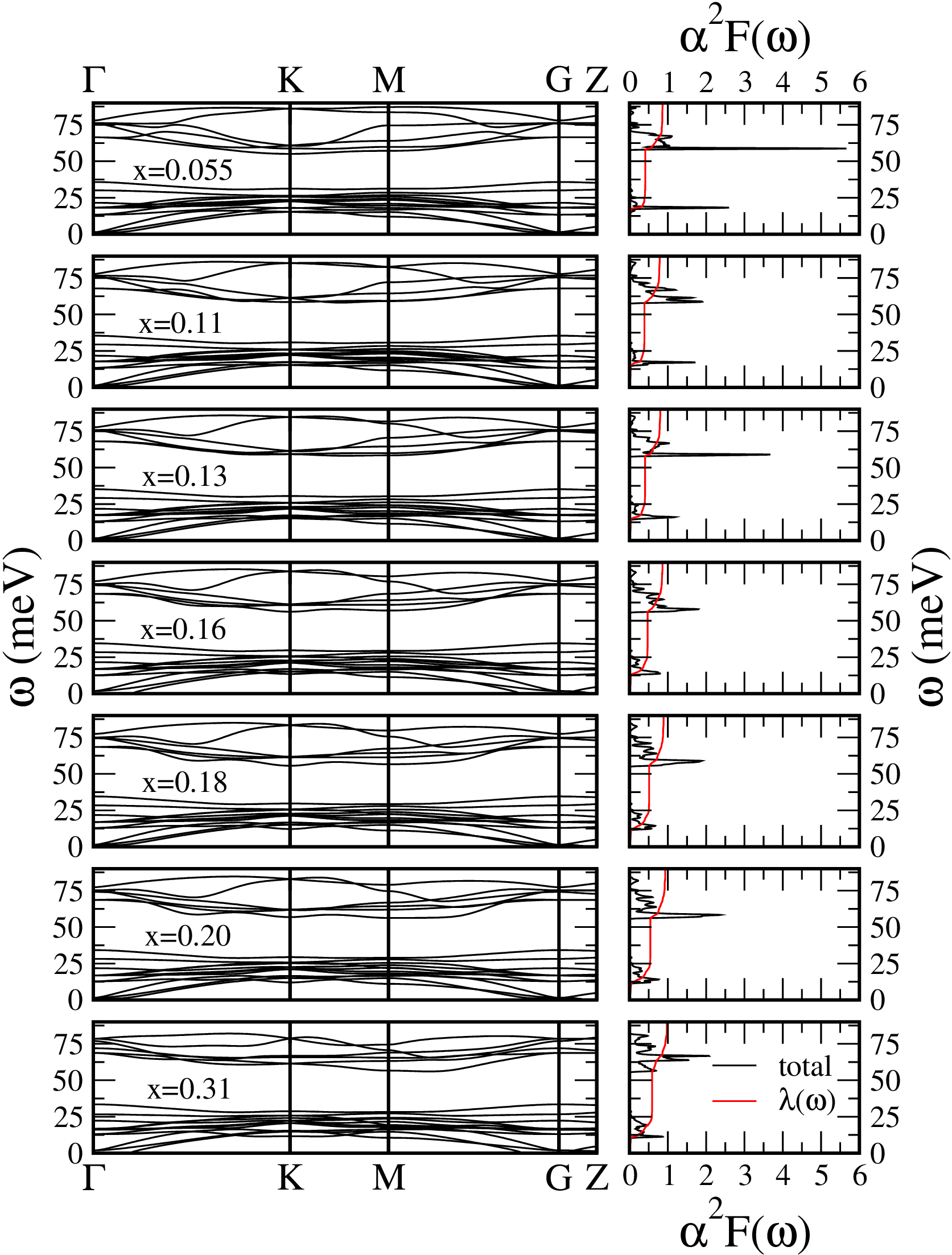}
        \caption{Left: Phonon dispersion of Li$_x$HfNCl as a function of doping.
        Right: Total Eliashberg function $\alpha^2F(\omega)$
        and electron-phonon coupling $\lambda(\omega)$.}
        \label{fig:phon}
\end{figure*}

In this Appendix, we present the phonon dispersion of Li$_x$HfNCl for all doping values.
The left panels of Fig. \ref{fig:phon} show the phonon dispersion with increasing doping.
Similarly the right panels show the corresponding Eliashberg function $\alpha^2 F(\omega)$
and the electron-phonon coupling $\lambda(\omega)$. 
In all cases, there are two distinct peaks of $\alpha^2 F(\omega)$, and consequently 
an increase in the $\lambda(\omega)$.

\clearpage

\bibliography{PaperBetul}

\end{document}